\documentclass[aps,prx,reprint,superscriptaddress,twocolumn,showpacs]{revtex4-1}

\usepackage{amsmath}
\usepackage{amssymb}
\usepackage{float}
\usepackage{dcolumn}
\usepackage{bm}
\usepackage{epsf}
\usepackage{graphicx}
\usepackage{amsfonts}
\usepackage{sidecap}
\sidecaptionvpos{figure}{c}
\usepackage{color}

\usepackage[colorlinks=true,citecolor=blue,linkcolor=blue,urlcolor=blue]{hyperref}

\newcommand{\done}{$3d_{x^2-y^2}${ }}
\newcommand{\dzz}{$3d_{z^2}${ }}
\newcommand{\dxy}{$3d_{xy/yz}${ }}
\newcommand{\NdNiO}{NdNiO$_{2}$}
\newcommand{\LaNiO}{LaNiO$_{2}$}
\newcommand{\f}{4\textit{f}{ }}
\newcommand{\dt}{3\textit{d}{ }}
\newcommand{\p}{2\textit{p}{ }}
\newcommand{\eg}{\textit{$e_g$}{ }}

\newcommand{\CA}{$C$-AFM{ }}
\newcommand{\GA}{$G$-AFM{ }}
\newcommand{\AM}{$A$-AFM{ }}

\newcommand{\JH}{J_\mathrm{H}}
\newcommand{\diff}{\mathrm{d}}

\definecolor{amaranth}{rgb}{0.9, 0.17, 0.31}

\begin{document}

\title{$f$-electron and magnetic ordering effects in nickelates \LaNiO~and \NdNiO: remarkable role of the cuprate-like \done band}

\author{Ruiqi~Zhang}\email{rzhang16@tulane.edu}
\affiliation{Department of Physics and Engineering Physics, Tulane University, New Orleans, LA 70118, USA}

\author{Christopher~Lane}
\affiliation{Theoretical Division, Los Alamos National Laboratory, Los Alamos, New Mexico 87545, USA}
\affiliation{Center for Integrated Nanotechnologies, Los Alamos National Laboratory, Los Alamos, New Mexico 87545, USA}

\author{Bahadur Singh}
\affiliation{Department of Condensed Matter Physics and Materials Science,
Tata Institute of Fundamental Research, Colaba, Mumbai 400005, India.}

\author{Johannes Nokelainen}
\affiliation{Department of Physics, School of Engineering Science, LUT University, FI-53850 Lappeenranta, Finland}
\affiliation{Department of Physics, Northeastern University, Boston, MA 02115, USA}

\author{Bernardo~Barbiellini}
\affiliation{Department of Physics, School of Engineering Science, LUT University, FI-53850 Lappeenranta, Finland}
\affiliation{Department of Physics, Northeastern University, Boston, MA 02115, USA}

\author{Robert S. Markiewicz}
\affiliation{Department of Physics, Northeastern University, Boston, MA 02115, USA}

\author{Arun~Bansil}\email{ar.bansil@neu.edu}
\affiliation{Department of Physics, Northeastern University, Boston, MA 02115, USA}

\author{Jianwei~Sun}\email{jsun@tulane.edu}
\affiliation{Department of Physics and Engineering Physics, Tulane University, New Orleans, LA 70118, USA}

\begin{abstract}
$\newline$

{Recent discovery of superconductivity in the doped infinite-layer nickelates has renewed interest in understanding the nature of high-temperature superconductivity more generally. The low-energy electronic structure of the parent compound NdNiO$_{2}$, the role of electronic correlations in driving superconductivity, and the possible relationship betweeen the cuprates and the nickelates are still open questions. Here, by comparing LaNiO$_2$ and NdNiO$_2$ systematically within a parameter free density functional framework, all-electron first-principles framework, we reveal the role Nd \f electrons in shaping the ground state of pristine NdNiO$_2$. Strong similarities are found between the electronic structures of \LaNiO~and \NdNiO, except for the effects of the 4$f$-electrons. Hybridization between the Nd \f and Ni 3$d$ orbitals is shown to significantly modify the Fermi surfaces of various magnetic states. In contrast, the competition between the magnetically ordered phases depends mainly on the gaps in the Ni $d_{x2-y2}$ band, so that the ground state in LaNiO$_2$ and NdNiO$_2$ turns out to be striking similarity to that of the cuprates. The $d-p$ band-splitting is found to be much larger while the intralayer 3$d$ ion-exchange coupling is smaller in the nickelates compared to the cuprates. Our estimated value of the on-site Hubbard $U$ is similar to that in the cuprates, but the value of the Hund's coupling $J_H$ is found to be sensitive to the Nd magnetic moment. The exchange coupling $J$ in \NdNiO~is only half as large as in the curpates, which may explain why $T_c$ in the nickelates is half as large as the cuprates.}

\end{abstract}

\maketitle

{\textbf{\small\color{amaranth}Introduction} }

Since the discovery of high-T$ _{c }$ superconductivity (HTSC) in the lanthanum-based cuprates in 1986~\cite{Bednorz1986}, understanding the mechanism of HTSC has drawn intense interest~\cite{Anderson2007,Lee2006,Anderson2004,Norman2011,Keimer2015}. Despite vigorous efforts, however, many unanswered questions still remain and a clear consensus on the underlying mechanism of HTSC has not been reached. A promising route in this connection is to find superconducting analogs of the cuprates which could provide new clues to the origin of HTSC. One such candidate materials is the perovskite nickel oxides. Specifically, the infinite-layer NdNiO$_{2}$ compound holds great promise since it exhibits an intrinsic 3$d^9$–filling much like the cuprates, although challenges of crystal growth have presented problems for undertaking systematic investigations of this material.

Recently, superconductivity in the hole-doped infinite-layer nickelate NdNiO$_{2}$ at 9$ \sim $15K has been reported in thin film samples grown on SrTiO$_3$~\cite{ Li2019a,Zeng2020a,Gu2020a,Zhang2020d}, although superconductivity in bulk NdNiO$_{2}$ has not been observed~\cite{BiXia}. These results~\cite{Li2019a} have reinvigorated interest in searching for the microscopic mechanism of HTSC and stimulated many new open questions ~\cite{Sawatzky2019, BiXia,Norman2020,Wu2020,Zhang2020a,Choi2020,Zhang2020d,Jiang2020,Botana2020,Karp2020,Li2019}. Notably, superconductivity is present both in hole-doped NdNiO$_{2}$~\cite{Li2019a} and PrNiO$_{2}$~\cite{Osada2020}, but absent in LaNiO$_2$~\cite{Li2019a}. This suggests that the Nd (Pr) $f$-electrons are not merely spectators, but possibly participate in the emergence of superconductivity. Interestingly, initial reports showed metallic behavior in pristine \LaNiO~and \NdNiO~\cite{Zhang2020d,Zeng2020a} with no sign of long range magnetic order, persisting down to low temperatures~\cite{Hayward2003,BiXia}, calling into question the role of Mott physics in HTSC. However, two recent transport studies~\cite{Zhang2020d,Zeng2020a} report the presence of a weak insulating phase in pristine \NdNiO, which could in part be a signature of short-range magnetic fluctuations due to the intrinsic off-stoichiometry produced by the inhomogeneous oxygen deintercalation crystal-growth process~\cite{Li2019a}.

In order to address these questions, a variety of theoretical studies have been performed, employing density functional theory (DFT)~\cite{Wu2020,Nomura2019,Jiang2019,Botana2020,Zhang2020c}, `beyond' DFT methods, such as DFT+U~\cite{Gu2020,Botana2020,Liu2020}, quasiparticle GW~\cite{Olevano2020}, dynamical mean-field theory (DMFT)~\cite{Lechermann2020,Ryee2020,Leonov2020,Gu2020}, and model Hamiltonians~\cite{Zhang2020a,Gu2020,Hu2019,Jiang2020} that have been constructed to understand the low-energy physics. The bulk of these studies focus on the NiO$_{2}$ plane, finding differences in quantities such as the $d-p$ orbital splitting as compared to the  cuprates~\cite{Botana2020} in accord with experimental reports~\cite{Hepting2020}. However, by focusing on the NiO$_{2}$ plane neglects the effects of the $f$-electrons on the electronic and magnetic structure, despite the presence of superconductivity in Nd and Pr based compounds but not in La~\cite{Li2019a,Osada2020}. The active role of the $f$-electrons is also suggested by a Kondo-like logarithmic temperature dependence of the resistivity and the Hall coefficient at low temperatures~\cite{Li2019a}, and other recent experiments demonstrating strong similarities between the infinite-layer nickelates and rare earth intermetallics~\cite{Hepting2020}, although a recent study attributes this strange behavior to the Nd 5$ d $ orbitals~\cite{Zhang2020a}. A few electronic structure studies utilizing the DFT+U~\cite{Choi2020} and Heyd-Scuseria-Ernzerhof (HSE) hybrid functional~\cite{Jiang2019} approaches have considered the $f$-electrons and find significant hybridization between the Nd \f and Ni 3$d$ orbitals near the chemical potential along with a possible ferromagnetic order. But, these calculations neglected the effect of spin-orbit coupling (SOC) crucial to capture the correct $f$-band splittings and required the introduction and fine tuning of external {\it ad hoc} parameters such as the Hubbard $U$ and the exact-exchange admixture, which limit their predictive power~\cite{Pokharel2020}. 

In this article, we present a systematic study of the electronic and magnetic structures of both \LaNiO~and \NdNiO~using the strongly-constrained-appropriately-normed (SCAN) density functional~\cite{Sun2015} with spin-orbit coupling to examine effects of the $f$-electron physics. The SCAN functional has a proven track record of accurately modeling many correlated materials families including the cuprates~\cite{Lane2018,Zhang2020,Furness2018,Nokelainen2020,Lane2020a}, iridates~\cite{Lane2020}, and ABO$_3$ materials~\cite{Varignon2019}. In particular, SCAN accurately predicts the $f$-band splitting in SmB$ _{6}$ in good accord with experimental values~\cite{ZhangR2020}. We consider several magnetic phases, whose energy and ordering are found to be quite similar for \LaNiO~and \NdNiO, reflecting their sensitivity to the opening of magnetic gaps in the Ni $d_{x2-y2}$ band. Dispersion of this band is quite similar to the corresponding band in cuprates, as is the order of the resulting magnetic phases, with an antiferromagnetic state having the lowest energy. In line with this, the estimated values of the Hubbard $U$ are close to those commonly found in the cuprates, while Hund $J_H$ varies for different Nd magnetic sublattice. In contrast, we find the the intralayer nearest-neighbor exchange coupling $J$ approximately one half as large as that in La$ _{2} $CuO$ _{4} $~\cite{Lane2018}. Lastly, the 4$f$-electrons play an important role in modifying the Fermi surfaces. We find the charge transfer energy between Ni \dt and O \p orbitals is large and does not change much with magnetic order. These latter findings suggest distinct physics in nickelates, which could be confirmed by further experiments.

$ \newline $
{\textbf{\small\color{amaranth}Results and Discussion}

\textbf{Crystal and magnetic structures:} Figure~\ref{fig:fig1} shows the crystal structure of \LaNiO~and \NdNiO~in the $P4/mmm$ symmetry~\cite{Hayward2003}, where NiO$_2$ planes are sandwiched together with La or Nd spacer layers. In the NiO$_2$ planes the Ni sites are surrounded by four O atoms in square-planar coordination. While the non-magnetic (NM) and ferromagnetic (FM) phases can be calculated in the primitive cell [Figs.\,\ref{fig:fig1} (a) and \ref{fig:fig1} (b)], the remaining three antiferromagnetic (AFM) phases require distinct supercells. Specifically, we use a $\sqrt2\times\sqrt2\times1$, $ \sqrt2\times\sqrt2\times2\ $, and $ 1\times1\times2 $ supercell for the $ C $-type AFM ($ C $-AFM), $ G $-type AFM ($ G $-AFM), and $ A $-type AFM ($ A $-AFM) orders, respectively, as shown in  Figs.\,\ref{fig:fig1} (c)-(e). In the $C$-AFM phase, the intralayer coupling in both Nd and Ni layers is AFM, whereas the interlayer is FM coupled. In the $G$-AFM phase both the intra- and interlayer coupling are AFM. In contrast, the $A$-AFM phase displays an intralayer FM coupling with an AFM interlayer coupling for both Nd and Ni sublattices. We note that the coupling between Ni and Nd nearest neighbors is frustrated in the $A$-AFM and $G$-AFM configurations. 

\begin{figure}[ht]
	\begin{center}
		\includegraphics[width=0.48\textwidth]{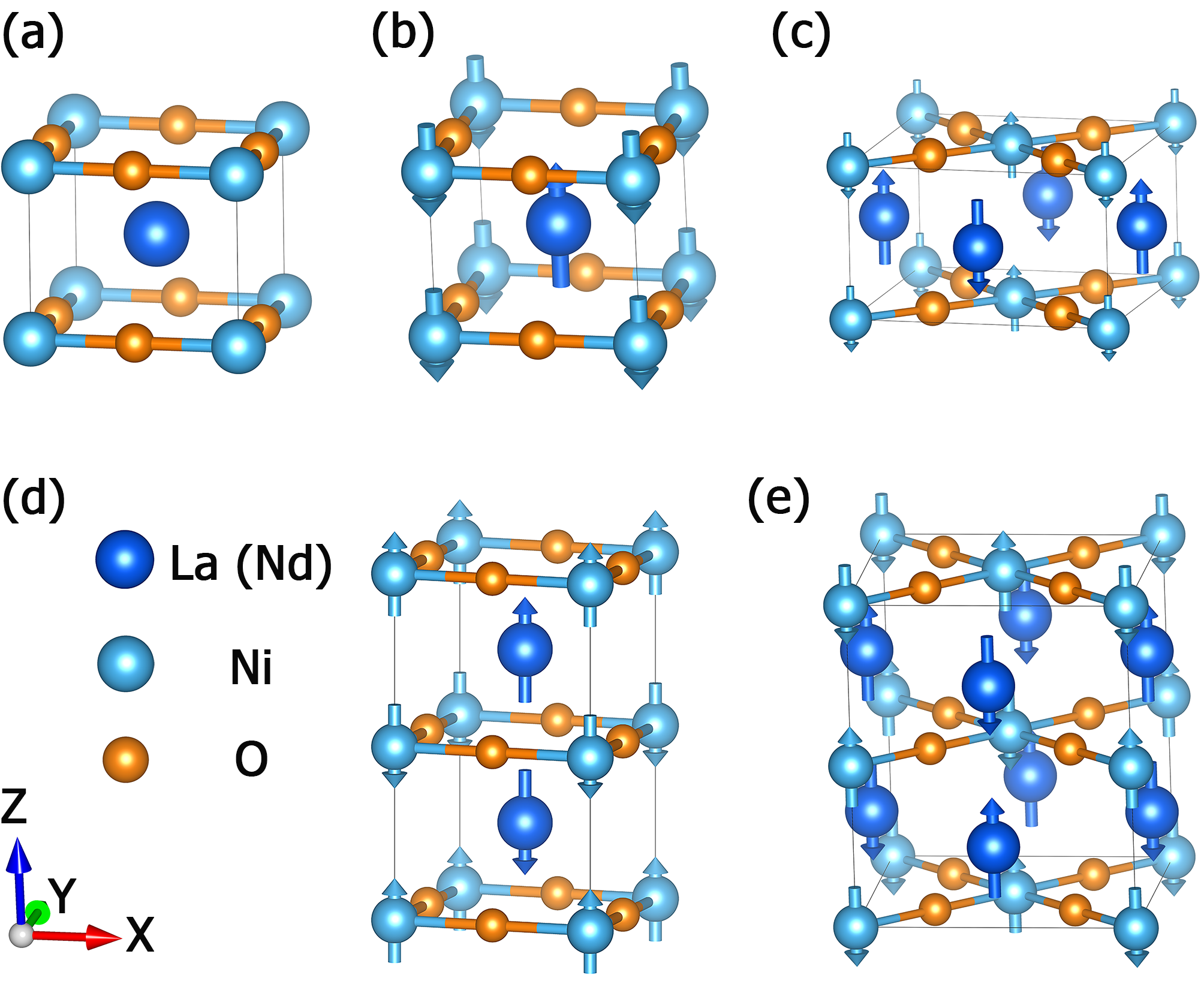}
	\end{center}
	\caption{(a) Non-magnetic (NM), (b) ferromagnetic (FM), (c) $C$-type antiferromagnetic ($C$-AFM),(d) $A$-
type antiferromagnetic ($A$-AFM) and (e) $G$-type antiferromagnetic ($G$-AFM) phase. The blue, gray, and yellow balls represent La (Nd), Ni and O atoms, respectively. The blue and gray arrows denote different magnetic moment directions. Note that there is no local magnetic moment on La.}
	
	\label{fig:fig1}
\end{figure}

Table \ref{Table1} gives our theoretically predicted total energies, lattice constants, and spin magnetic moments for various magnetic phases of \LaNiO~and \NdNiO. For \LaNiO, the \CA phase is the most stable, with the $G$-AFM, $A$-AFM, and NM phases lying at higher energies. Upon optimizing the crystal structure we find the predicted lattice parameters for the magnetic phases in good accord with the experimental values, while those from the NM phase are quite further away. Moreover, the lattice constant and energy of the FM ($C$-AFM) phase are almost same as in the \AM ($G$-AFM) phase, suggesting that interlayer coupling in \LaNiO~ is very weak. Finally, our theoretically predicted local nickel magnetic moment is $\sim$1.0 $\mu_B$ irrespective of the magnetic configuration. 

Our results are in contrast to those found with GGA+U~\cite{Botana2020,Lee2004} ($U=3$) which predict a much smaller energy separation between the magnetic configurations along with reduced magnetic moments of $\sim$0.7 $\mu_B$ and $\sim$0.5 $\mu_B$ in the \CA and FM phases, respectively. The reduced moment values could possibly be due to SOC effects being neglected, but this result is still surprising since a significant Hubbard $U$ was introduced on the Ni sites.

\begin{table}[bht]
	\setlength{\belowcaptionskip}{0.5cm}
	\renewcommand\arraystretch{1.5}
	\caption{ Comparison of various theoretically predicted properties for different magnetic phases of \LaNiO~and NdNiO$_2$. The lattice constants are for the primitive cell, and the energies are relative to reference $C$-AFM phase. The magnetic moments in these phases have small variations in magnitude among the different sites of Nd and Ni, so the results given here are the average values. M$ _{Nd}$ and M$ _{Ni}$  represent the local magnetic moment per Nd and Ni atom, respectively.} 
	\label{Table1}
	\centering
	\tabcolsep 0.15cm
	\begin{tabular}{c|ccc|c|c|c}\hline\hline
	    \multicolumn{7}{c}{LaNiO$_2$}\\  \hline
		Phases & \multicolumn{3}{c|}{Lattice Constant (\AA{})} & Energy &\multicolumn{2}{c}{$M _{Ni}$} \\
		& a \quad &b\quad  &c \quad & (meV/f.u.) & \multicolumn{2}{c}{($\mu $B)}  \\ \hline
		NM	& 3.896 \quad &3.896 \quad  &3.384 \quad & +404  &\multicolumn{2}{c}{0}   \\
		$C$-AFM	& 3.947 \quad &3.947\quad  &3.347 \quad & 0 & \multicolumn{2}{c}{1.02}  \\ 
		$G$-AFM	& 3.947 \quad &3.947 \quad  &3.345 \quad & +21  & \multicolumn{2}{c}{0.96}  \\
		FM	& 3.942 \quad &3.942 \quad  &3.354 \quad & +62 & \multicolumn{2}{c}{0.98}  \\
		$A$-AFM	& 3.941 \quad &3.941\quad  &3.358 \quad & +60  & \multicolumn{2}{c}{1.00}  \\ 
		Exp.~\cite{Kaneko2009}	& 3.959 \quad &3.959 \quad  &3.375 \quad & - & \multicolumn{2}{c}{-} \\
		\hline\hline
		\multicolumn{7}{c}{NdNiO$_2$}\\  \hline
		Phases & \multicolumn{3}{c|}{Lattice Constant (\AA{})} & Energy & $M _{Nd} $ &$M _{Ni}$ \\
		& a \quad &b\quad  &c \quad & (meV/f.u) & ($ \mu $B) & ($ \mu $B)  \\ \hline
		NM	& 3.866 \quad &3.866 \quad  &3.222 \quad & +3840 & 0 & 0   \\
		$C$-AFM	& 3.910 \quad &3.910\quad  &3.239 \quad & 0 & 2.91 & 1.04  \\
		$G$-AFM	& 3.911 \quad &3.911 \quad  &3.239 \quad & +134 & 2.91 & 0.94  \\
		FM	& 3.874 \quad &3.874 \quad  &3.298 \quad & +65 & 3.04 & 0.94  \\
		$A$-AFM	& 3.887 \quad &3.887\quad  &3.269 \quad & +141 & 2.95 & 0.94  \\ 
		Exp.~\cite{Li2019}	& 3.914 \quad &3.914 \quad  &3.239 \quad & - & - & - \\
		\hline\hline
	\end{tabular}
\end{table}

We now compare our  \LaNiO~results to the corresponding properties of NdNiO$_2$ (see Table \ref{Table1}). For both compounds the $C$-AFM configuration is the ground state, with almost the same energy separation with the FM phase. However, the energy differences between the other AFM phases are much larger in NdNiO$_2$ than those in \LaNiO, with the NM phase lying at 3840 meV/f.u. The larger energy differences in NdNiO$_2$ between the FM (\GA) and \AM (\CA) phases suggests that the interlayer exchange coupling is strongly affected by the presence of the local magnetic moment of Nd. Furthermore, like \LaNiO, the lattice constants of NdNiO$_2$ in the $C$- and $G$-AFM phases are extremely close to the experimental values compared to those obtained in the NM, FM, and \AM arrangements. The stabilization of \CA order over that of \GA appears to be due to the frustration between Nd and Ni magnetic moments, clearly illustrating the importance of the Nd $f$-electrons in the calculation.

Table~\ref{Table1} shows the theoretically predicted local magnetic moments of Nd and Ni to be $\sim$3.0 $\mu_B$ and $\sim$1.0 $\mu_B$, respectively. This suggests that the electron configuration of Nd and Ni are [Xe] 4$f^{3}$ and 3$d^9$, respectively. By breaking the Ni magnetic moment down into the various orbital contributions, we find 0.75 $\mu_B$ and 0.25 $\mu_B$ on the $d_{x^2-y^2}$ and $d_{z^2}$ orbitals, respectively, where the $t_{2g}$ states have negligible moments. A number of theoretical studies in the literature have reported magnetic ordering and the associated local moments~\cite{Lee2004,Liu2020,Choi2020a,Choi2020} with strong spin fluctuations possibly playing a key role in cooper pairing~\cite{Leonov2020,Choi2020a,Chang2019}. Specifically, the GGA finds the local moment on Ni to be significantly reduced, yielding $\sim$0.52 and  $\sim$0.35 $\mu_B$ in the FM and AFM phases, respectively~\cite{Choi2020}. 
Where the SCAN values are only recovered when a large Hubbard U of 8 (5) eV for Nd (Ni) is assumed~\cite{Choi2020}. Moreover, a previous calculation utilizing the SCAN functional~\cite{Zhang2020c} surprisingly finds reduced Ni magnetic moment values of 0.76 $\mu_B$ \cite{Zhang2020c}, but these calculations neglect Nd \f electrons and SOC effects. Lastly, the HSE06 hybrid functional finds stabilized moments of  $\sim$3.03 $\mu_B$ and $\sim$0.89 $\mu_B$ on  Nd and Ni sites, respectively, for the FM phase~\cite{Jiang2019}, similar to the SCAN values. Overall, the SCAN predictions are closely aligned with the expected $d$ and $f$-filling of NdNiO$_2$ without any fine-tuning.

\textbf{Electronic structure of NM Phase:}  Figures~\ref{fig:fig2} (a) and \ref{fig:fig2} (b) show the theoretically obtained  band structures and density of states (DOS) for \LaNiO~and~\NdNiO~ in the non-magnetic (NM) phase, with the various orbital-resolved atomic site projections overlaid. For \LaNiO, Fig.~\ref{fig:fig2} (a),  we find two distinct bands crossing the Fermi level: one of nearly pure Ni-\done character, and the other composed of Ni ($3d_{z^2}$, $3d_{xy/yz}$) and La 5$d$ orbitals. The latter band produces a 3D spherical electron-like Fermi surface at $\Gamma$ and $A$ [Fig.~\ref{fig:fig2} (c)], whereas the former generates a large slightly warped quasi-2D cylindrical Fermi surface similar to the cuprates. These findings are consistent with previous LDA~\cite{Botana2020}, GGA~\cite{Jiang2019}, DFT+U~\cite{Lee2004}, and LDA+DMFT ~\cite{Ryee2020} studies.  We note that the \done band bears a striking resemblance to the corresponding band in cuprates~\cite{Lane2018}, except for a shift of the  VHS energy between $\Gamma$ and $Z$ planes in the Brillouin zone. The shifting of the VHS from below to above the Fermi level along $k_z$ is directly reflected in the Fermi surface transitioning from being open to closed in the $\Gamma$ and $Z$ planes, respectively [Fig.~\ref{fig:fig2} (c)].

\begin{figure}[ht]
	\begin{center}
		\includegraphics[width=0.48\textwidth]{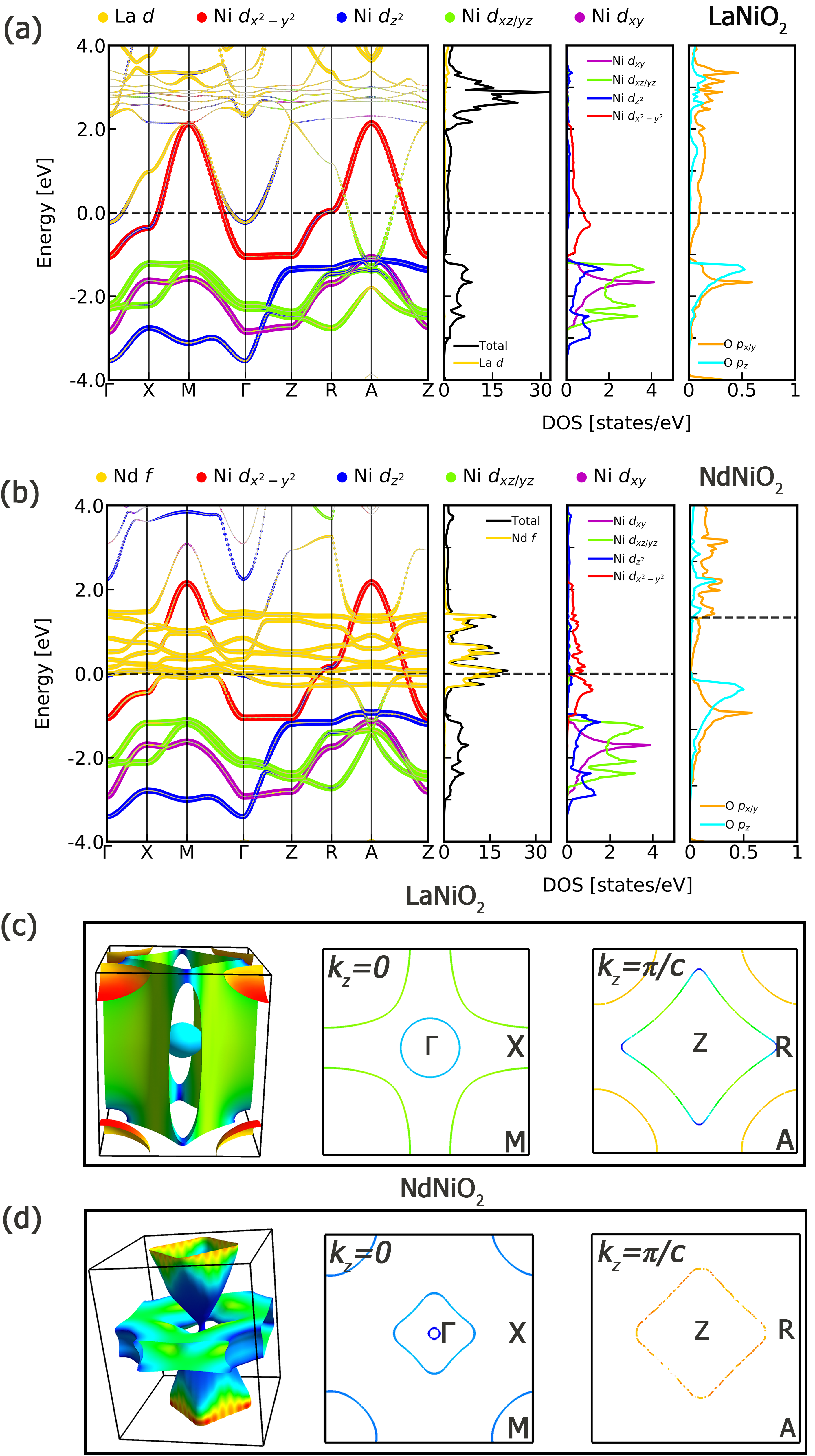}
	\end{center}
	\caption{Orbital projected electronic structures of NM (a) \LaNiO~and (b) NdNiO$_2$. (c) The calculated Fermi surface of \LaNiO~and the projection of Fermi surface on $k _{z} $ = 0 and $\pi/c$ plane. (d) Same as (c) but for NdNiO$_2$. Red, green, and blue indicate the maximum, middle, and minimum Fermi velocity, respectively.}
	
	\label{fig:fig2}
\end{figure}

On comparing \LaNiO~with NdNiO$_2$ [Fig.~\ref{fig:fig2} (b)], we find all the Ni bands to be relatively unchanged, except for significant hybridization between the Nd 4$f$-electrons and Ni \done orbitals. Interestingly, this similarity between the Ni 3$d$ dispersions in \LaNiO~ and  \NdNiO~  persists across all magnetic phases studied [Figs.\,\ref{fig:fig3}--\ref{fig:fig6}]. The significant hybridization between Ni 3$d$ and Nd 4$f$ levels can give rise to self-doping effects and induce Kondo physics~\cite{Lechermann2020,Liu2020,Nomura2019,Zhang2020c,Gu2020}. Such hybridizations also radically alter the Fermi surfaces. Figure~\ref{fig:fig2} (d) displays a double-goblet-like hole pocket along the $\Gamma$-Z direction, with a very narrow stem at $\Gamma$. Moreover, a large and complex hole Fermi surface appears surrounding $\Gamma$ near the $ k_{z}=0 $ plane, along with the formation of electron pockets near $M$. We further note that the narrow neck of the goblet Fermi surface at $\Gamma$ suggests that the system is close to a Fermi surface topological transition where the goblet splits into Fermi pockets centered on the $Z$ point [Fig.~\ref{fig:fig2} (d)].

\textbf{Electronic structure of C-AFM Phase:} Figures \ref{fig:fig3} (a) and \ref{fig:fig3} (b) present the unfolded theoretical electronic structure of \LaNiO~and \NdNiO~ in the \CA phase. Similar to the cuprates, the $C$-AFM order is stabilized by opening a 2 eV band gap in the $d_{x^2-y^2}$ dominated band. However, unlike the cuprates, 5$d$ and 4$f$ states fill the gap maintaining the metallic nature of these compounds.  For example, the states near the Fermi level in \LaNiO~are mainly governed by the Ni \dzz and La 5$d$ orbitals [Fig.~\ref{fig:fig3} (a)]. Moreover, an extremely flat band is found pinned near the Fermi level along the $Z-R-A-Z$ line, originating from the La 5$d$ and Ni \dzz hybridized orbitals. A similar flat band is found in \NdNiO, along with a second flat band along $Z-R-A$, stemming from Nd 4$f$––Ni \dxy hybridization. These flat band features have also been observed by Choi et. al~\cite{Choi2020a} where a large U was used to push the Nd 4$f$ states away from the Fermi level. These flat-band features also produce highly anisotropic Fermi surfaces near the k$_{z}$=$\pi/c$ plane for both \LaNiO~ and \NdNiO~[Fig.\,\ref{fig:fig3} (c) and (d)]. For \NdNiO~ there are two additional heavy electron pockets ellipsoidal in shape [Fig.\,\ref{fig:fig3}(d)], which are induced by strong Nd \f and Ni 3d$_{z^2}$ hybridization.

\begin{figure}[ht]
	\begin{center}
		\includegraphics[width=0.48\textwidth]{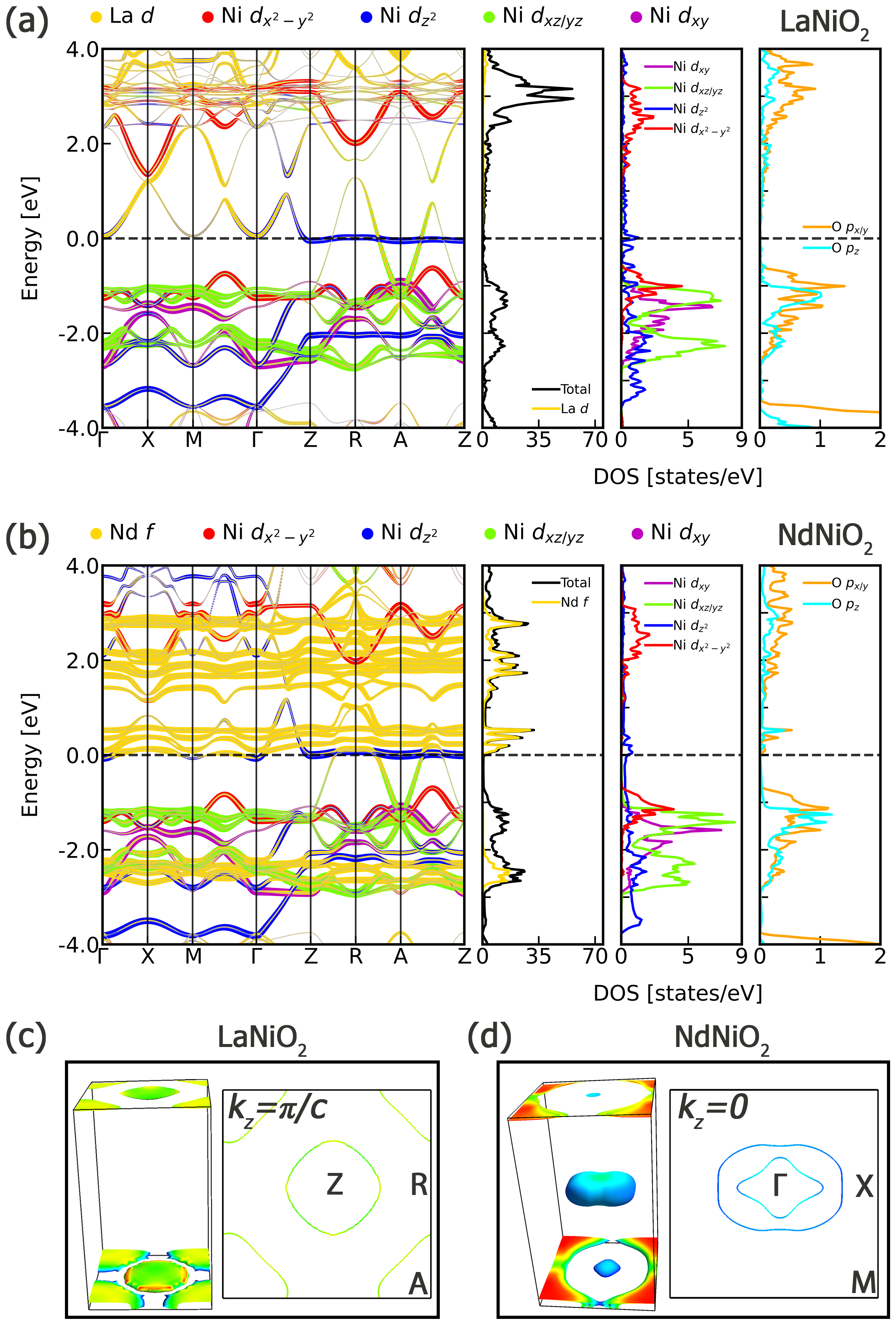}
	\end{center}
	\caption{Orbital projected electronic structures of \CA phase of (a) \LaNiO~ and (b) NdNiO$_2$. The band structures are unfolded to the primitive $1 \times 1 \times 1$ BZ. (c) The calculated Fermi surface of \LaNiO~ and the projection of Fermi surface on $k _{z} $ = $\pi/c$ plane. (d) Same as (c) but for NdNiO$_2$. Red, green, and blue indicate the maximum, middle, and minimum Fermi velocity, respectively.}
	
	\label{fig:fig3}
\end{figure}

\textbf{Electronic structure of G-AFM Phase:} While the $C$-AFM phase is found to be the ground state, it is important to study other low-lying phases in correlated quantum materials that could contribute to various intertwined orders~\cite{Lane2018,Lane2020,Zhang2020,ZhangR2020,Varignon2019}. Figure~\ref{fig:fig4} (a) and \ref{fig:fig4}(b) show the electronic structures of \LaNiO~and \NdNiO~ in the \GA phase. The \GA magnetic phase exhibits AFM coupling between the intra- and interlayer magnetic sites, in contrast to $C$-AFM where the interlayer coupling is FM.  Interestingly, here we find no $f$-bands near the Fermi level for \NdNiO, making the Fermi surfaces in  \LaNiO~and \NdNiO~ equivalent. While both \GA and \CA phases are dominated by the splitting of the $d_{x2-y2}$-band,  and both have a region of suppressed DOS within $\sim$0.7 eV of the Fermi level,  they differ in that a \dzz band is above the low DOS region in the \CA phase, but below it in the \GA phase, causing the Fermi energy to shift by $\sim$0.7 eV.

\begin{figure}[ht]
	\begin{center}
		\includegraphics[width=0.48\textwidth]{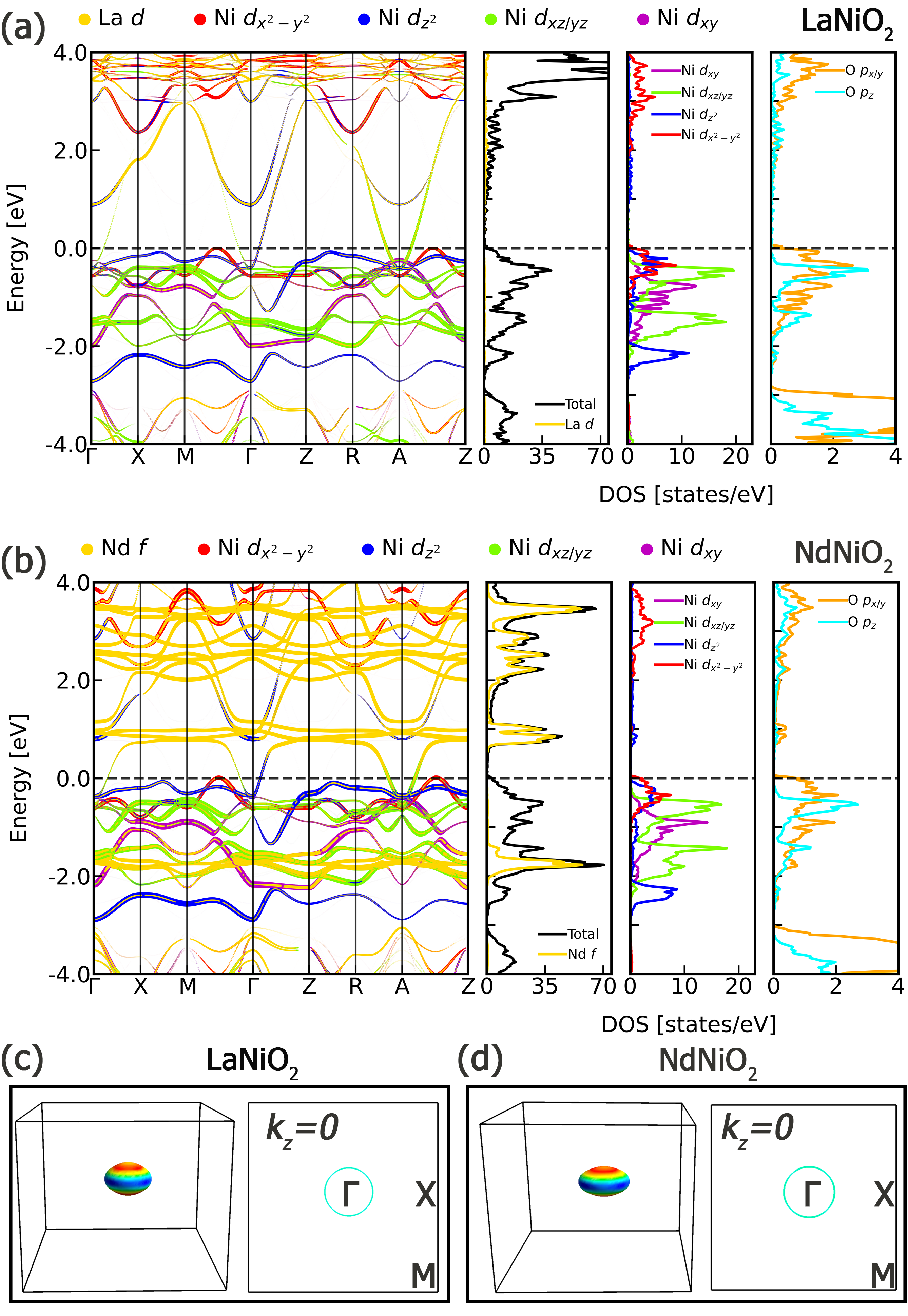}
	\end{center}
	\caption{Orbital projected electronic structures of \GA phase of (a) \LaNiO and (b) NdNiO$_2$.The band structures are unfolded to the primitive $1 \times 1 \times 1$ BZ. (c) The calculated Fermi surface of \LaNiO and the projection of Fermi surface on $k _{z} $ = 0 plane. (d) Same as (c) but for NdNiO$_2$. Red, green, and blue indicate the maximum, middle, and minimum Fermi velocity, respectively.}
	
	\label{fig:fig4}
\end{figure}

\textbf{Electronic structure of FM Phase:} Figures~\ref{fig:fig5} (a) and \ref{fig:fig5} (b) present the electronic structure of \LaNiO~and \NdNiO~ in the FM phase. Although the spin moments of Nd and Ni were initialized in the same direction, the system selfconsists into a ferrimagnetic configuration [Fig.~\ref{fig:fig1} (b)]. Compared with the NM phase [Fig.~\ref{fig:fig2}], the Ni \done and \dzz bands are spin split, due to the spin-polarization in FM phase. In this case the Fermi surface is composed of a Ni \done hole pocket (red) at the $M$ point and an electron pocket from the hybridization between Ni \dzz and Nd 4$f$ orbitals at $\Gamma$ point (blue) [Fig.~\ref{fig:fig5} (b)]. Interestingly, the majority spins in the Ni \done and \dzz bands point in opposite directions for a given atom.  Figs.\,\ref{fig:fig5} (c) and (d) show that the Fermi surfaces of these two compounds are quite similar, except that the $\Gamma$ point electron pocket in Fig.~\ref{fig:fig5} (c) has grown `propellers' in Fig.~\ref{fig:fig5} (d), which is due to hybridization between Ni \dzz and Nd 4$f$ orbitals. We also find a 2D Fermi sheet centered at the $M$ point extending in the $k_z$ direction, produced by hybridization between the Ni \dxy and Nd \f orbitals. Finally, there is an A-centered hole pocket generated by the Ni \done band.

\begin{figure}[ht]
	\begin{center}
		\includegraphics[width=0.48\textwidth]{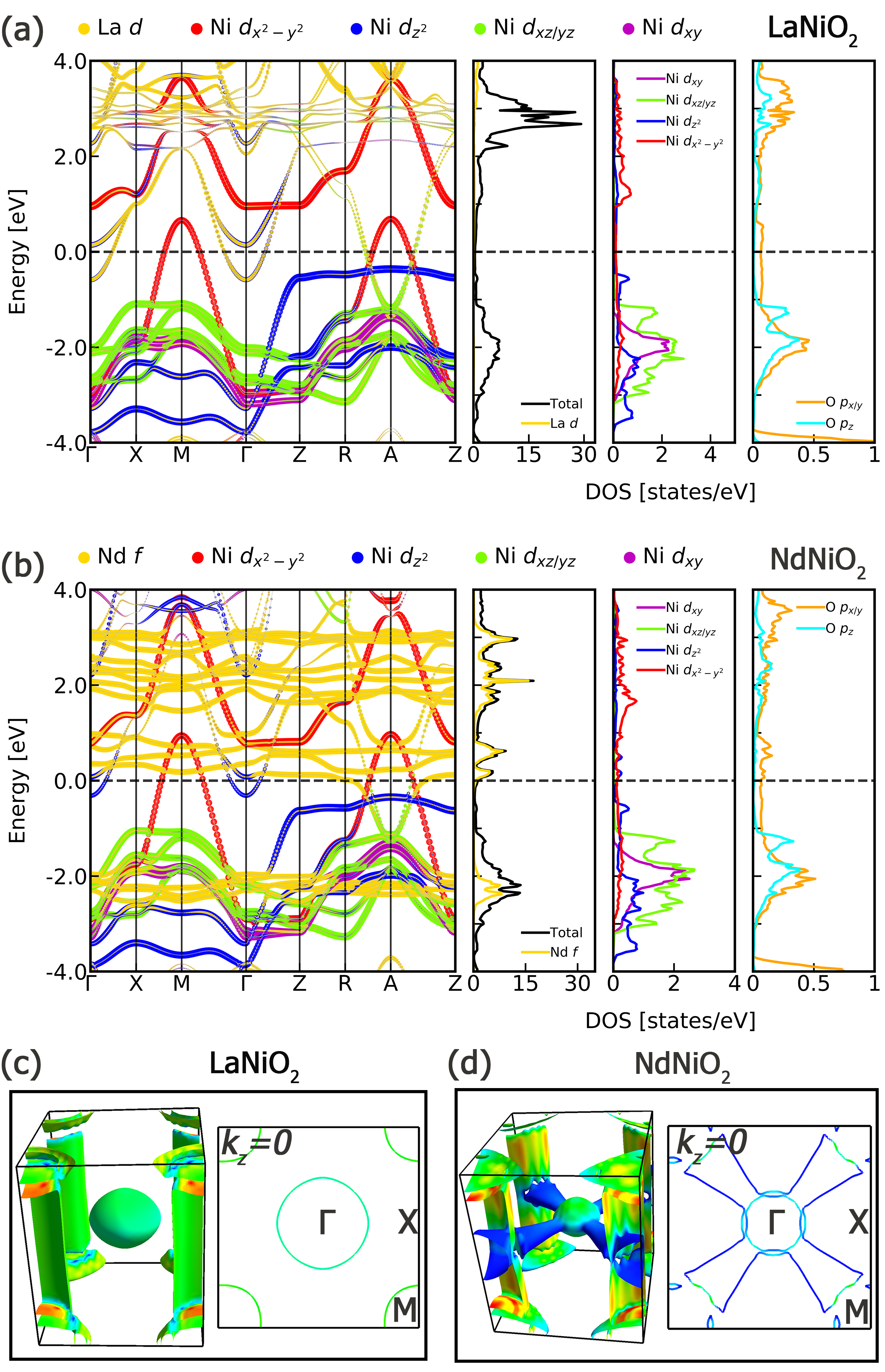}
	\end{center}
	\caption{Orbital projected electronic structures of FM phase of (a) \LaNiO and (b) NdNiO$_2$. (c) The calculated Fermi surface of \LaNiO and the projection of Fermi surface on $k _{z} $ = 0 plane. (d) Same as (c) but for NdNiO$_2$. Red, green, and blue indicate the maximum, middle, and minimum Fermi velocity, respectively.}
	
	\label{fig:fig5}
\end{figure}

\textbf{Electronic structure of A-AFM Phase:} Figures~\ref{fig:fig6} (a) and \ref{fig:fig6} (b) display the unfolded electronic band structure and density of states of \LaNiO~and \NdNiO~in the \AM phase. Since the NiO$_2$ layers are ferromagnetically ordered, the resulting band-splitting is quite similar to the FM phase [Fig.~\ref{fig:fig5}]. Figure~\ref{fig:fig6} (a), shows that the bands near the Fermi level in \LaNiO~are mainly of Ni $3d_{x^2-y^2}$, and hybridized Ni \dzz and La 5$d$ character. However, in NdNiO$_2$, we find the main low-lying states near the Fermi level to originate from Nd 4$f$ states hybridizing with Ni \done and \dzz orbitals. Notably, around -1 eV we see a strong mixing between the Ni \done and \dzz orbitals in both \LaNiO~and \NdNiO~ along $\Gamma-X$ and $R-A$ directions, which are absent in the NM and FM phases. Such a strong `orbital-mixing' effect could make the physics in the nickelates quite different from the cuprates.  

\begin{figure}[ht]
	\begin{center}
		\includegraphics[width=0.48\textwidth]{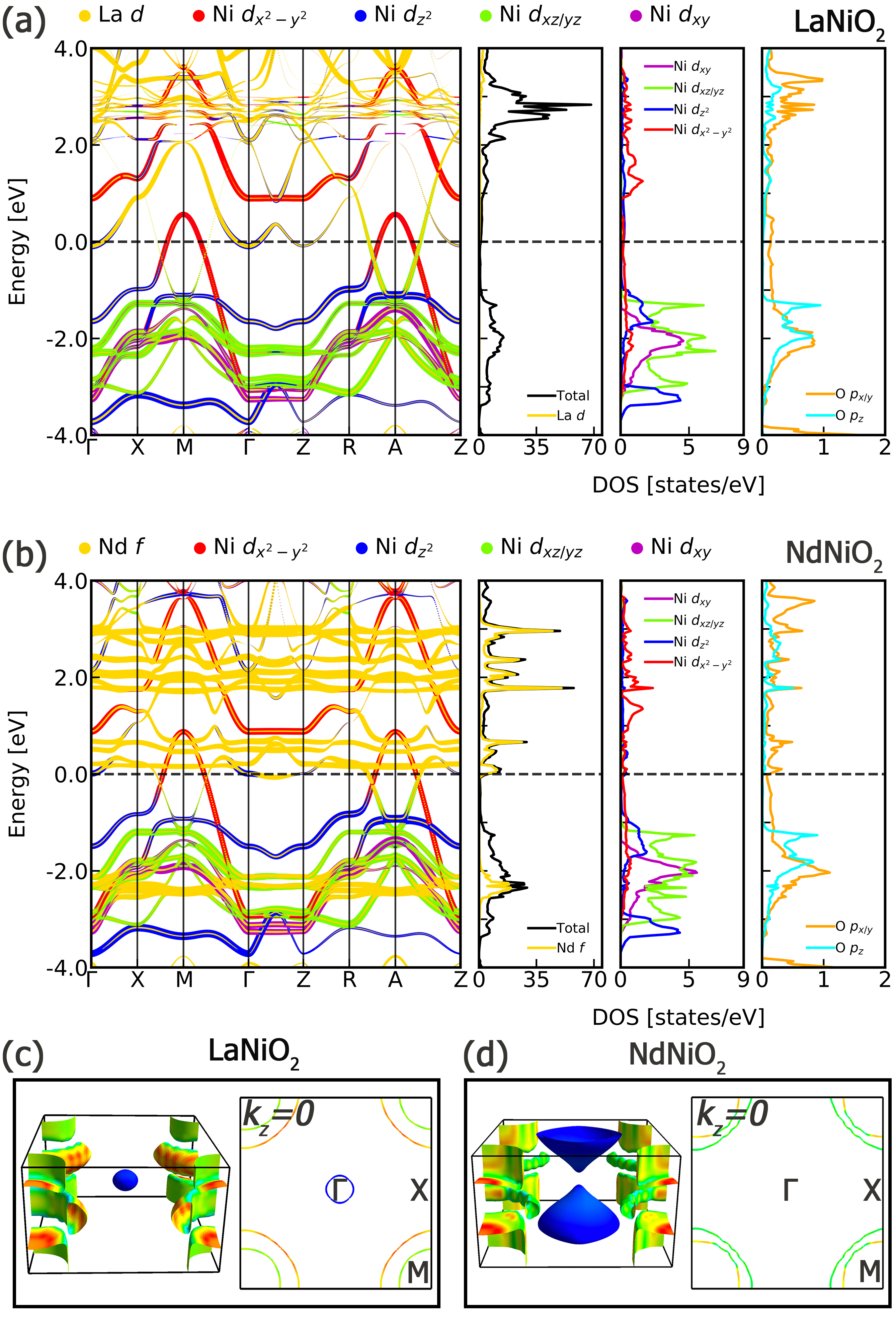}
	\end{center}
	\caption{Orbital projected electronic structures of \AM phase of (a) \LaNiO~and (b) NdNiO$_2$. The band structures are unfolded to the primitive $1 \times 1 \times 1$ BZ. (c) The calculated Fermi surface of \LaNiO~and the projection of Fermi surface on the $k _{z} $ = 0 plane. (d) Same as (c) but for NdNiO$_2$. Red, green, and blue indicate the maximum, middle, and minimum Fermi velocity, respectively.}
	\label{fig:fig6}
\end{figure}

From Figs.~\ref{fig:fig6} (c) and \ref{fig:fig6} (d), the Fermi surfaces of \LaNiO~and \NdNiO~fall into two categories. (i) The $M-A$ direction in both materials is similar to the corresponding FM Fermi surfaces, except for the splitting along the z-axis due to the AFM stacking of adjacent FM layers.  This effect leads to the appearance of two pockets near the M point in $\Gamma$-plane. (ii) The $\Gamma-Z$ direction is more reminiscent of the NM case, with a Ni \dzz electron pocket at the $\Gamma$ point in \LaNiO, while in \NdNiO, the goblet Fermi surface of the NM phase has split into Z-centered pockets.  The blue color of these features (low Fermi velocity) suggests strong $f$-electron mixing.

\textbf{$f$-electron dispersions:} Based on the preceding comparison of \LaNiO~and \NdNiO, we find the stabilization of the various magnetic phases of \NdNiO~to be mainly driven by Ni $d$-electrons, with the $f$-electrons playing only a minor role. In contrast, the Fermi surfaces are strongly affected by the Nd 4$f$-electrons, exhibiting strong mixing with Ni 3$d$ orbitals. While the $f$-electrons form one large cluster in the NM phase, they split into three subbands once the Nd atoms become polarized with slight shifts depending on the magnetic order consistent with our recent work on SmB$_6$~\cite{ZhangR2020}.

\textbf{Comparisons with cuprates:} Table \ref{Table2} gives the calculated $\Delta_{dp}$,  $\Delta_{\eg}$, U, and $\JH$ for the various magnetic phases of \LaNiO~and \NdNiO, along with the corresponding values for the cuprates for comparison. The values of $\Delta_{dp}$ for \LaNiO~phases range from 1.58 (FM) to 2.08 eV (\GA), while the corresponding values for \NdNiO~span 1.82 to 2.84 eV. To illustrate the charge-transfer gap further and compare to \CA of the infinite-layer cuprate CaCuO$_2$ the partial density of states for the Ni (Cu) 3$d$ and O 2$p$ orbitals is shown in Figs.~\ref{fig:fig7} (a) and  (c). Here, the O 2$p$ band-center is clearly lower than the center of gravity of the Ni 3$d$ states by $\sim$2 eV in \NdNiO, whereas the O 2$p$ levels are strongly hybridized with Cu \done orbitals near the Fermi level in CaCuO$_2$. To quantify this, we find $\Delta_{dp}$ for  CaCuO$_2$ in the \CA phase to be 0.19 eV, which is significantly smaller than nickel-based compounds. Additionally, we estimated $\Delta_{dp}$ for the single-layer La$_2$CuO$_4$ in the \GA phase to be 0.6 eV, still much smaller than nickelates. Based on these results, \LaNiO~and \NdNiO~ are closer to the Mott-Hubbard limit rather than charge-transfer case based on the Zaanen–Sawatzky–Allen classification scheme~\cite{Zaanen1985}. 

\begin{figure}[ht]
	\begin{center}
		\includegraphics[width=0.48\textwidth]{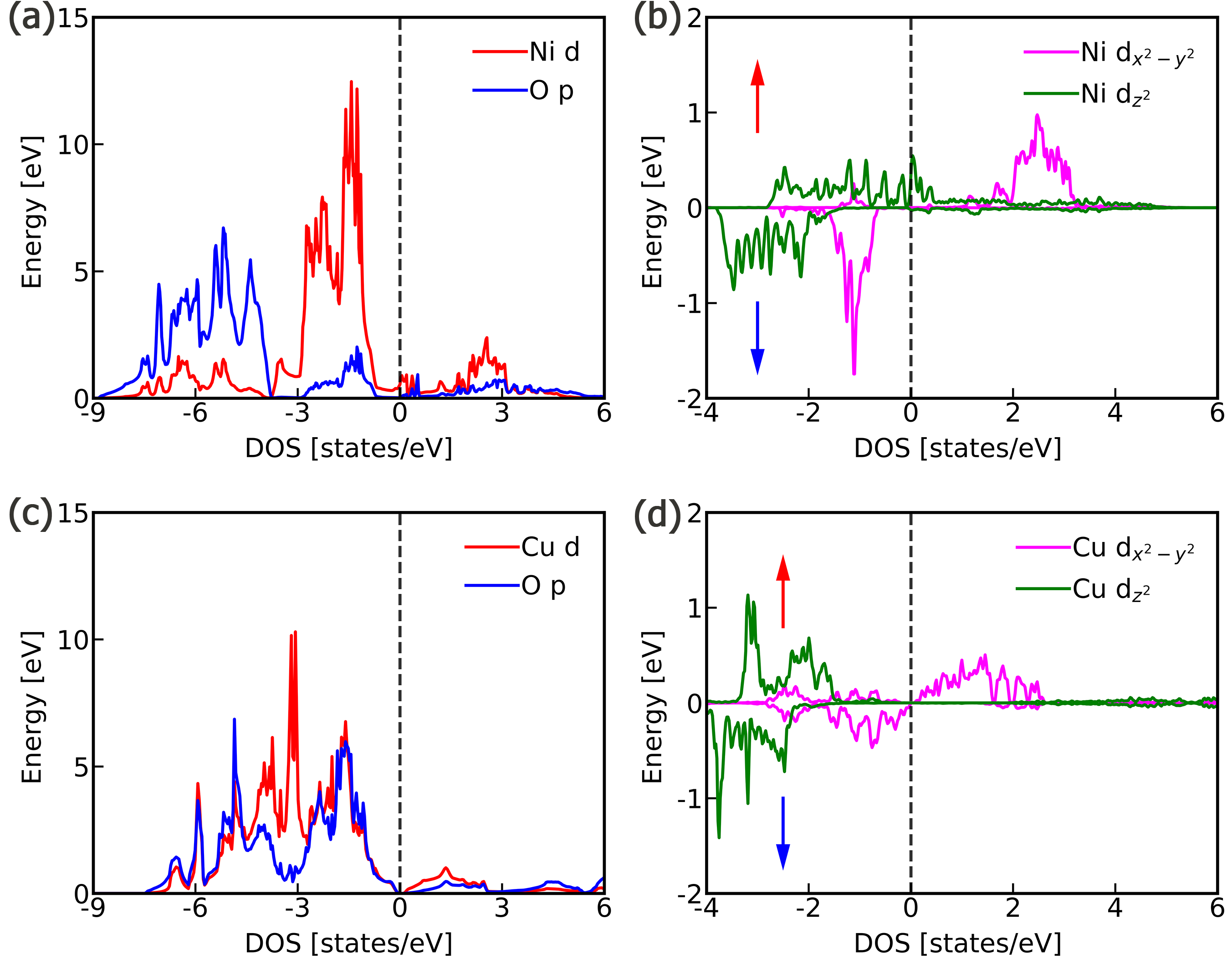}
	\end{center}
	\caption{(a) Partial densities of states in the \CA phase of NdNiO$_2$, where red and blue lines represent the Ni \d and O 2$p$ orbitals, respectively. (b) Single nickel-site-resolved partial densities of states in the \CA  phase of NdNiO$_2$, where the magenta and green lines represent the Ni \done and \dzz orbitals, respectively. And, the red and blue arrows represent the spin up and spin down channels, respectively. (c) and (d) same as (a) and (b), except that these panels refer to the \CA phase of CaCuO$_2$.}
	
	\label{fig:fig7}
\end{figure}

Interestingly, values of $\Delta_{\eg}$ for the various magnetic phases of \LaNiO~and \NdNiO~are very close to 2 eV. For example, the $ \Delta_{\eg}$ for \LaNiO~ in the NM phase is 1.93 eV. Our estimate of $\Delta_{\eg}$ is consistent with the value of 1.95 eV obtained in Ref.~\onlinecite{Botana2020}. The similarity of $ \Delta_{\eg}$ values across the infinite-layer nickelates suggests that the Nd 4$f$ electrons play a very limited role in splitting the Ni 3$d$ levels. Additionally, we estimate the value of $\Delta_{eg}$ for \CA CaCuO$_{2}$ and \GA La$_2$CuO$_4$ to be close to those obtained from the nickelates [Table~\ref{Table2}].  This information also can be read from  Figs.~\ref{fig:fig7} (b) and (d), where $\eg$ orbitals are quite splitting for both \CA \NdNiO~and CaCuO$_2$. 

To gauge the strength of correlations on the nickel site we calculate the Hubbard $U$ and Hund's coupling $\JH$ for \LaNiO~and \NdNiO~ in various magnetic arrangements according to Eq.(1)-(3). We find the on-site potential $U$ for \LaNiO~and \NdNiO~to be almost the same and very close to values obtained for CaCuO$_{2}$ and La$_2$CuO$_4$, suggesting strong electron interactions in nickelates. Note that these estimated U values are also consistent with recent works~\cite{Nomura2019,Sakakibara2019}.
The behavior of $\JH$ is more subtle. In general, $\JH$ for nickelates and cuprates are very similar, whereas $J_H$s for \NdNiO~are larger in the \AM and \GA phases, suggesting that $\JH$ is highly sensitive to the interlayer coupling.

\begin{table}[bht]
	\setlength{\belowcaptionskip}{0.5cm}
	\renewcommand\arraystretch{1.5}
	\caption{Comparison of properties of different phases of nickelates and cuprates. $\Delta_{dp}$ and $ \Delta_{\eg} $ represent the splitting of the metallic (Ni and Cu) \dt and O \p bands, and the splitting of tradition metal ions (Ni and Cu) \eg bands, respectively. U and $\JH$ are on-site Hubbard potential and Hund’s coupling, respectively.} 
	\label{Table2}
	\centering
	\tabcolsep 0.15cm
	\begin{tabular}{ccccc}\hline\hline
	    \multicolumn{5}{c}{LaNiO$_2$}\\  \hline
		Phases & $\Delta_{dp}$ (eV) & $\Delta_{\eg}$ (eV) &U (eV) &$\JH$ (eV) \\
		NM	& 1.91 \quad & 1.93 \quad  &-- \quad & --   \\
		FM	& 1.58 \quad & 2.23 \quad  &4.88 \quad & 1.01   \\
		$A$-AFM	 & 1.57 \quad & 1.93 \quad  &4.99 \quad & 1.14   \\
		$C$-AFM	& 1.68 \quad & 2.02 \quad  &5.21 \quad & 1.11   \\
		$G$-AFM	& 2.08 \quad & 2.09 \quad  &5.70 \quad & 1.45   \\
		\hline\hline
		\multicolumn{5}{c}{NdNiO$_2$}\\  \hline
		Phases & $\Delta_{dp} $ & $\Delta_{\eg}$ &U (eV) &$\JH$ (eV) \\
		NM	& 2.84 \quad & 1.94 \quad  &-- \quad & --   \\
		FM	& 1.82 \quad & 2.04 \quad  &5.08 \quad & 1.23   \\
		$A$-AFM	 & 1.80 \quad & 2.00 \quad  &4.99 \quad & 2.14   \\
		$C$-AFM	& 2.59 \quad & 2.11 \quad  &5.27 \quad & 1.00  \\
		$G$-AFM	& 2.25 \quad & 2.13 \quad  &5.78 \quad & 1.94   \\
		\hline\hline
		\multicolumn{5}{c}{CaCuO$_2$}\\  \hline
		Phases & $\Delta_{dp} $ & $\Delta_{\eg}$ &U (eV) &$\JH$ (eV) \\
		$C$-AFM	& 0.19 \quad & 2.65 \quad  &5.35 \quad & 1.29  \\
		\hline\hline
		\multicolumn{5}{c}{La$_2$CuO$_4$}\\  \hline
		Phases & $\Delta_{dp} $ & $\Delta_{\eg}$ &U (eV) &$\JH$ (eV) \\
		$G$-AFM	& 0.6 \quad & 1.23 \quad  &4.85~\cite{Lane2018} \quad & 1.25~\cite{Lane2018}  \\
		\hline\hline
	\end{tabular}
\end{table}

Despite there being no clear theoretical description of the mechanism of HTSC, the view that spin-fluctuations play a central role in determining the physical properties of the cuprates has been gaining increasing support. Furthermore, in this picture, the exchange coupling strength is a good descriptor of the robustness of superconductivity.

In order to determine the strength of the exchange coupling, we map the total energies of the AFM and FM phases onto those of the nearest-neighbor spin-1/2 Heisenberg Hamiltonian in the mean-field approximation~\cite{su1999crystal,Lane2018}. The Heisenberg Hamiltonian gives a reasonable description of the low-lying excitations for La$_2$CuO$_4$, and thus a good estimate of the Heisenberg exchange parameter $J$~\cite{Kastner1998}, and we expect this also to be the case for the infinite-layer nickelates. In the mean-field limit, the difference in total energies of the FM and AFM phases is

\begin{align}
    \Delta E=E_{AFM}-E_{FM}=JNZS^2
\end{align}

where $N$ is the total number of magnetic moments, $S$ is the spin on each site, and $Z$ is the coordination number. Since the in-plane interactions within the Ni-O planes in La(Nd)NiO$_2$ are much stronger than the interplanar interactions, we can take $Z = 4$. Since we normalize to one formula unit, $N = 1$. Using the total energies for the FM and $C$-AFM states obtained from our first-principles computations we find $J$ to be 62 and 65 meV for \LaNiO~and \NdNiO, respectively. These exchange parameters are half as large as those in  La$ _{2} $CuO$ _{4} $~\cite{Lane2018} and larger than those estimated in Ref.~\cite{Jiang2020}. The small $J$ is consistent with the finding that the AFM gap is twice as large as in curpates, which in turn may be related to the larger value of Ni magnetic moments, which may be related to smaller Ni-$d$ and O-$p$ hybridization (larger $\Delta_{dp}$).

\begin{table}[bht]
    \centering
	\setlength{\belowcaptionskip}{0.5 cm}
	\renewcommand\arraystretch{2.0}
	\caption{Comparison of the intralayer exchange coupling between the two nearest Ni or Cu atoms for \LaNiO, \NdNiO, and La$ _{2} $CuO$ _{4} $.} 
	\label{Table3}
	\tabcolsep 0.3cm
	\begin{tabular}{cccc}\hline\hline
	    	& \LaNiO  & \NdNiO   & La$_{2}$CuO$_{4}$\\ \hline
		$J$ (meV)	& 62  & 65  & 138~\cite{Lane2018} \\
		\hline\hline
	\end{tabular}
\end{table}

Superconductivity in the cuprates evolves out of a Mott insulator~\cite{Lee2006}, whereas in the iron pnictides superconductivity emerges out of a metallic state~\cite{Norman2008} with strong local magnetic fluctuations. Is magnetic order necessary for $d$-electron high-$T_c$ superconductivity?  The new Ni-based superconductors appear to be a counterexample. However, both Ni and Nd sites typically display significant magnetic moments with considerable evidence of magnetic fluctuations or short-range order~\cite{Choi2020a,Leonov2020}. Moreover, the undoped nickelates are not ordinary metals, but weak insulators~\cite{Zhang2020d,Zeng2020a}. In our previous SCAN-based studies of other correlated materials, we found many low-energy magnetic phases indicative of prominent magnetic fluctuations~\cite{Zhang2020,Lane2018,ZhangR2020}.  In the nickelates, our study of the various AFM orders finds $\sim$0.7 eV pseudogap (regions of low DOS) near the Fermi level, which could explain the weak insulating behavior.  
In Figure~\ref{fig:fig8}, we compare the AFM gaps in the ground state structures of the cuprates (a) and the nickelates (b). We find the gap is about twice as large in the nickelates as compared to the cuprates~\cite{Lane2018}, which may explain why $J$ is only half as large in the former. 

\begin{figure}[ht]
	\begin{center}
		\includegraphics[width=0.48\textwidth]{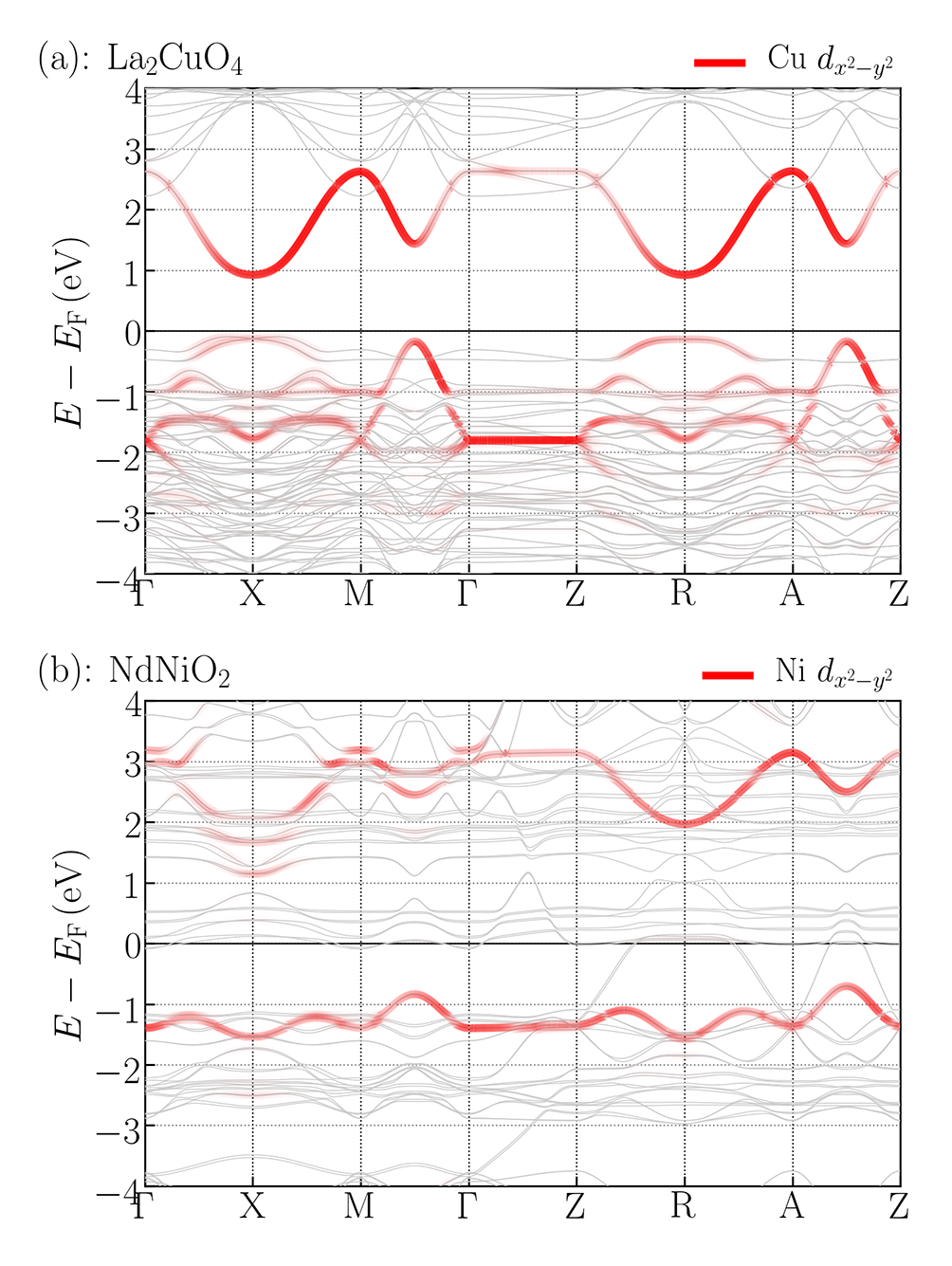}
	\end{center}
	\caption{(a) Cu \done band for \GA La$_2$CuO$_4$. (b) Ni \done~band for \CA NdNiO$_2$. Band structures have been unfolded into the $1 \times 1 \times 1$ primitive BZ.}
	
	\label{fig:fig8}
\end{figure}

Hence, we find great similarity between the cuprates and nickelates, both in the dispersion of the NM $d_{x^2-y^2}$ band, and in the resulting magnetic orders, with the $f$-electrons playing little role in the magnetic transitions despite significantly modifying the Fermi surfaces.  In principle, the fact that $J$ is only half as large in nickelates as in cuprates could tell us something about why the $T_c$ dome is only half as large in nickelates.  On the other hand, so far superconductivity has been found in two rare-earth substituted nickelates~\cite{Li2019a,Osada2020}, but not the parent La-based compound, suggesting a more significant role for $f$-electrons.  An interesting possibility is that the $f$-electrons could lead to heavy-fermion physics (flat bands) not present in cuprates.

\textbf{Conclusion:} We examine in-depth the role of $f$-electrons and magnetic ordering effects in \LaNiO~and NdNiO$_2$ within a parameter-free, all-electron first-principles framework. The magnetic orders in the nickelates are found to be very similar to those in the cuprates in that the transitions are driven by the gapping of the $d_{x^2-y^2}$ band. We find a reduced $J$ value in the nickelate compared to the cuprates, however, which could explain the weaker superconductivity in the nickelates. While the \f electrons play little role in the nickelate magnetism, they substantially modify Fermi surfaces for various magnetic states. Our study further reveals the importance of fluctuating magnetic order in correlated materials~\cite{Zhang2020}.

$ \newline $

{\textbf{\small\color{amaranth}Methods }

All calculations were performed by using the pseudopotential projector-augmented wave method~\cite{Kresse1999} as implemented in the Vienna {\it ab initio} simulation package (VASP)~\cite{Kresse1993,Kresse1996}. A high-energy cutoff of 520 eV was used to truncate the plane-wave basis set. The exchange-correlation effects were treated using the SCAN~\cite{Sun2015} meta-GGA scheme. Spin-orbit coupling effects were included self-consistently. The crystal structures and ionic positions were fully optimized with a force convergence criterion of 0.01  eV/\AA{}  for each atom along with a total energy tolerance of 10$ ^{-5} $ eV. The Fermi surface was obtained with the FermiSurfer code~\cite{Kawamura2019}. The unfolded band structures including orbital characters are extracted from the supercell pseudo-wavefunction  calculation~\cite{Popescu2012}, which has been implemented based on the VaspBandUnfolding code~\cite{QZ2017}.

To facilitate comparison with the cuprates, we calculated two quantities: (1) charge-transfer energies between the Ni \dt and O \p orbitals $\Delta_{dp} = \varepsilon_d - \varepsilon_p$ and (2) the energy splitting of the two Ni \eg orbitals $\Delta_{\eg} = \varepsilon_{x^2-y^2} - \varepsilon_{z^2}$. Here,  $ \varepsilon_i $ refers to the band centers of the corresponding orbital $i$. Following previous works,~\cite{Botana2020,Jang2015} the band centers are defined as $ \varepsilon_i=\frac{\int{g_i(\varepsilon)\varepsilon d\varepsilon}}{\int{g_i(\varepsilon) d\varepsilon}}  $, where $ g_i(\varepsilon) $ refers to the partial-density-of-states (PDOS) associated with orbital $ i $. The integration range for $ \Delta_{dp} $ is set to cover the full bonding and antibonding bands~\cite{Jang2015}, whereas $ \Delta_{\eg} $ is obtained from an integral over the antibonding bands alone, using an energy window of -3.5 to 2 eV and -4 to 4 eV for the NM and magnetic phases, respectively.

To estimate the on-site Hubbard potential U and the Hund’s coupling J$ _{H} $, we follow the method developed by Lane \textit{et al.}~\cite{Lane2018}.  Using the site-projected orbital-resolved partial DOS $ g_{\mu\sigma} $, we determine the average spin-splitting of the $\mu$ levels as follows:

\begin{gather}
	\overline{E}_{\mu\sigma}
	=
	\int_W E g_{\mu\sigma}(E) \,\diff E,
	\\
	\overline{E}_{d_{x^2-y^2}\uparrow} - \overline{E}_{d_{x^2-y^2}\downarrow}
	=
	U (N_\uparrow-N_\downarrow),
	\\
	\overline{E}_{\mu\neq d_{x^2-y^2}\uparrow} - \overline{E}_{\mu\neq d_{x^2-y^2}\downarrow}
	=
	\JH(N_\uparrow - N_\downarrow),
\end{gather}

where $N_\uparrow$ ($N_\downarrow$) is the occupation of the spin-up (down) 
$d_{x^2-y^2}$ orbital and the integration is over the full bandwidth $W$.

$ \newline $
{\textbf{\small\color{amaranth}Data availability}

The data that support the findings of this study are available from the corresponding author upon reasonable request.

\bibliography{Ref}

$ \newline $
{\textbf{\small\color{amaranth}Acknowledgements}

The work at Tulane University was supported by the start-up funding from Tulane University, the Cypress Computational Cluster at Tulane, the Extreme Science and Engineering Discovery Environment (XSEDE), the DOE Energy Frontier Research Centers (development and applications of density functional theory): Center for the Computational Design of Functional Layered Materials (DE-SC0012575), the DOE, Office of Science, Basic Energy Sciences Grant DE-SC0019350, and the National Energy Research Scientific Computing Center. The work at Northeastern University was supported by the U.S. DOE, Office of Science, Basic Energy Sciences grant number DE-FG02-07ER46352 (core research) and benefited from Northeastern University’s Advanced Scientific Computation Center, the National Energy Research Scientific Computing Center supercomputing center (DOE grant number DEAC02-05CH11231), and support (testing the efficacy of new functionals in diversely bonded materials) from the DOE Energy Frontier Research Centers: Center for the Computational Design of Functional Layered Materials (DE-SC0012575). The work at Los Alamos National Laboratory was supported by the U.S. DOE NNSA under Contract No. 89233218CNA000001 and by the Center for Integrated Nanotechnologies, a DOE BES user facility, in partnership with the LANL Institutional Computing Program for computational resources. Additional support was provided by DOE Office of Basic Energy Sciences Program E3B5.\\

$ \newline $
{\textbf{\small\color{amaranth}Author contributions} }

R.Z., C.L., B. S. and J. N. performed computations and analyzed the data. B.B., R.S.M.,
A.B. and J.S. led the investigations, designed the computational approaches, provided
computational infrastructure and analyzed the results. All authors contributed to the
writing of the manuscript.

$ \newline $
{\textbf{\small\color{amaranth}Additional information} }

\textbf{Competing interests:} The authors declare no competing interests.


\end{document}